\documentclass{pnastwo}

\usepackage{subcaption, wrapfig, cite}
\newcommand{\nat}{{ Nature }}
\newcommand{\aap}{{Astron. \& Astrophys. }}
\newcommand{\aj}{{ Astron.~J. }}
\newcommand{\gca}{{Geochim. Cosmochem. Acta}} 
\newcommand{\grl}{{Geophys. Res. Letters}}
\newcommand{\apj}{{ Astrophys.~J. }}

\newcommand{\apjl}{{Astrophys.~J.~Letters }}
\newcommand{\apjs}{{Astrophys.~J.~Suppl. }}

\newcommand{\icarus}{{Icarus }}

\newcommand{\mnras}{{MNRAS }}

\newcommand{\ssr}{{Space Sci. Rev.}}

\url{www.pnas.org/cgi/doi/10.1073/pnas.0709640104}
\copyrightyear{2008}
\issuedate{Issue Date}
\volume{Volume}
\issuenumber{Issue Number}
\begin{document}

\title{Tracing the Ingredients for a Habitable Earth from Interstellar Space through Planet Formation
}

\author{Edwin A. Bergin\affil{1}{University of Michigan},
Geoffrey A. Blake\affil{2}{California Institute of Technology},
Fred J. Ciesla\affil{3}{University of Chicago}
Marc M. Hirschmann\affil{4}{University of Minnesota}
\and
Jie Li\affil{1}{University of Michigan}}

\contributor{Accepted by Proceedings of the National Academy of Sciences
of the United States of America}

%%%Newly updated.
%%% If significance statement need, then can use the below command otherwise just delete it.
\significancetext{With the rapid pace at which exoplanets are being discovered, many efforts have now been dedicated to identifying which planets are expected to have the ingredients necessary for the development of life.  In this work we explore the relative disposition of the essential elements carbon and nitrogen in each stage of star and planet formation, using the Earth and our solar system as grounding data. Our results suggest that planets like the Earth are readily supplied with these key elements, but their relative amounts on the surface and in the atmosphere will be highly variable.}

\maketitle

\begin{article}
\begin{abstract}
{  We use the C/N ratio as a monitor of the delivery of key
ingredients of life to nascent terrestrial worlds. Total elemental
C and N contents, and their ratio, are examined for the
interstellar medium, comets, chondritic meteorites and
terrestrial planets; we include an updated estimate for the
Bulk Silicate Earth (C/N = 49.0 $\pm$ 9.3). Using a kinetic  model of
disk chemistry, and the sublimation/condensation temperatures
of primitive molecules,
we suggest that organic ices and macro-molecular (refractory or carbonaceous dust) organic material are the likely
initial C and N carriers.
Chemical reactions in the disk can produce
nebular C/N ratios of $\sim$1$- $12, comparable to those of
comets and the low end estimated for planetesimals.
An increase of the C/N ratio is
traced between volatile-rich pristine bodies and larger
volatile-depleted objects subjected to
thermal/accretional metamorphism. The C/N ratios of the dominant
materials accreted to terrestrial planets should therefore be higher
than those seen in carbonaceous chondrites or comets.  During planetary
formation, we explore scenarios leading to further
volatile loss and associated C/N variations owing to core formation and
atmospheric escape.  Key processes include relative enrichment of
nitrogen in the atmosphere and preferential sequestration of carbon by
the core. The high C/N BSE ratio therefore is best satisfied by accretion
of thermally processed objects followed by large-scale atmospheric
loss. These two effects must be more profound if volatile sequestration
in the core is effective.  The stochastic nature of these processes
hints that the surface/atmospheric abundances of biosphere-essential
materials will likely be variable.}
\end{abstract}


%\abbreviations{SAM, self-assembled monolayer; OTS, octadecyltrichlorosilane}

\dropcap{T}he development of  a habitable world and a stable biosphere requires the delivery of biogenic elements of which carbon and nitrogen are crucial.  Carbon is the backbone for the chemistry of life and, 
in the form of CO$_2$, combines with water to provide the greenhouse needed for a habitable Earth.  Nitrogen is a key component of DNA, RNA, and proteins, while also present as the dominant constituent of our atmosphere.   The processes that supply these crucial ingredients remain poorly understood.  In interstellar space, C and N are abundant, but inherently volatile and so chiefly remain in the gas.  Thus, the terrestrial planets, which accrete primarily from rocks and ices, are fed from C- and N-depleted materials and are carbon and nitrogen poor compared to the nebular disk from which they descend \cite{Halliday2013, pontppvi}.    The carbon and nitrogen depletion of rocky bodies is a general phenomenon, observable not just in our solar system, but in the polluted atmospheres of white dwarf stars, which trace the composition of disrupted planetesimals \cite{Jura06, Gansicke12}.  
 This volatile poor state of terrestrial planets is partially imparted from the starting materials. However, further differential loss of C and N can occur due to parent body processes such as thermal metamorphism, core segregation, planetary outgassing, and atmospheric loss \cite{huss06,brearley_messii, Marty12, Dasgupta2013b, Roskosz13, Chi14, Tucker14}.

In this work we document the evolution in the relative concentrations of C and N from the 	 medium (ISM) through planetary assembly. We show that the C/N ratio contains clues regarding the formation of terrestrial planets and the delivery/fate of crucial volatile compounds. We first discuss the relative carbon and nitrogen inventories beginning with the ISM as it evolves to the protoplanetary disk, employing the perspectives of kinetic chemistry, our growing understanding regarding the composition of planet-forming disks, and volatile loss within thermally processed planetesimals. We then explore the evolution of C/N during geochemical differentiation of a young Earth-like planet, built up by accretion of planetesimals of a range of sizes. Key processes during this last stage of C/N evolution include sequestration into the metallic core and outgassing to the nascent atmosphere. It is of course, the latter that can provide ingredients for early environments and life, but early atmospheres are also prone to impact-driven escape, providing additional mechanisms for C and N processing and alteration.

\section{Carbon and Nitrogen Inventories}

Fig.~\ref{fig:CN} displays the carbon-to-nitrogen ratio observed in the Sun, Bulk Silicate Earth (BSE; defined in Supplementary Information), meteoritic samples, comets, and that inferred in the ISM. To provide an absolute reference frame, Fig.~\ref{fig:siabun} provides bulk C and N abundances normalized to Si.\footnote{Further details regarding this compilation, including technique where appropriate, are found in Supporting Information and Table S1.}
Key considerations include:

\vspace{-0.05in}
\begin{enumerate}
\item  
 In the ISM, significantly more carbon than nitrogen is incorporated into carbonaceous dust (e.g., Fig.~\ref{fig:siabun}), with a correspondingly lower C/N ratio ($<$solar) in the gas. The primary form of condensed carbon is uncertain, but models assume it is an aliphatic/aromatic hydrocarbon mixture 
\cite{Jones13, Chiar13}. 
%Using these numbers to compare to bulk cometary measurements it is clear that cometary ices are not entirely comprised of interstellar components, at least in terms of their refractory (dust) materials (e.g. Fig.~\ref{fig:siabun}). 
 Interstellar ices have elevated C/N ratios, with a median value of C/N $\sim$ 12 \cite{oberg11_c2d}.  
However, this does not account for N or N$_2$, which are infrared inactive.  Correcting for the presence of N$_2$ in ices (see Supplementary Information) leads to a C/N ratio for ices of $\sim$1.8.

\item Owing to low levels of observed nitrogen, particularly in ices (e.g., N$_2$), comets have elevated C/N ratios when compared to the Sun\cite{wte91}. 
%Thus, in the solar nebular disk nitrogen must have been in more volatile form when compared to carbon. 
Based on the relative disposition of C and N in the ISM this can be partially achieved as comets accrete ISM carbonaceous dust with low nitrogen content. 
In this light, Halley is inferred to have more N in solid dust than its gaseous coma
\cite{wte91}, which is also higher than estimates for interstellar dust (Fig.~\ref{fig:siabun}).
% perhaps hinting that additional sources of refractory nitrogen-bearing carbonaceous material (i.e. organics) were available in the solar nebular disk. 
There is also a sharp difference in the {\em absolute} C and N content of Halley when compared to the Sun-grazing comets, which have chondritic abundances.  However, their C/N ratios are comparable. 
Anhydrous Interplanetary Dust Particles (IDP), which are argued to be of cometary origin \cite{Messenger_messii}, are also characterized by a high carbon content, (C/Si)$_{\rm IDP}$ $\sim$ 2 \cite{Thomas93}.
%The abundance of both volatile element, relative to silicon, is about a factor of $\sim$50 deficient in the Sun-grazing comets than in Comet Halley. 

\item Carbonaceous, enstatite, and ordinary chondrites
have elevated C/N ratios compared to both the Sun and comets.  
 All meteorites are carbon-poor when compared to
the Sun, ISM dust, and comet Halley, but the key distinction is owing to
comparative nitrogen depletion in chondrites. Among meteorites,
carbonaceous, and enstatite chondrites have similar ratios, but ordinary
chondrites have higher and more variable C/N, with mean values closer to
the Bulk Silicate Earth.

 \item  
 For the Earth, both the surface reservoir and Bulk Silicate Earth
have elevated C/N ratios when compared to the Sun. The surface reservoir
is comparable to chondritic, owing to preferential volcanic degassing of
N$_2$, leading to the N-rich atmosphere.  There are two other estimates of the C/N ratio
for the Earth, C/N $\sim$ 53$\pm$20 \cite{Halliday2013} and 425$\pm$329 \cite{Marty12} (see Supplementary Information); in each case the estimated C/N ratio of the BSE is quite high.\footnote{Limited estimates for Mars and Venus are supplied in the Supplementary Information.}

\end{enumerate}
\vspace{-0.07in}

\begin{figure}[t]
\begin{center}
\centerline{\includegraphics[angle=0, width=.5\textwidth]{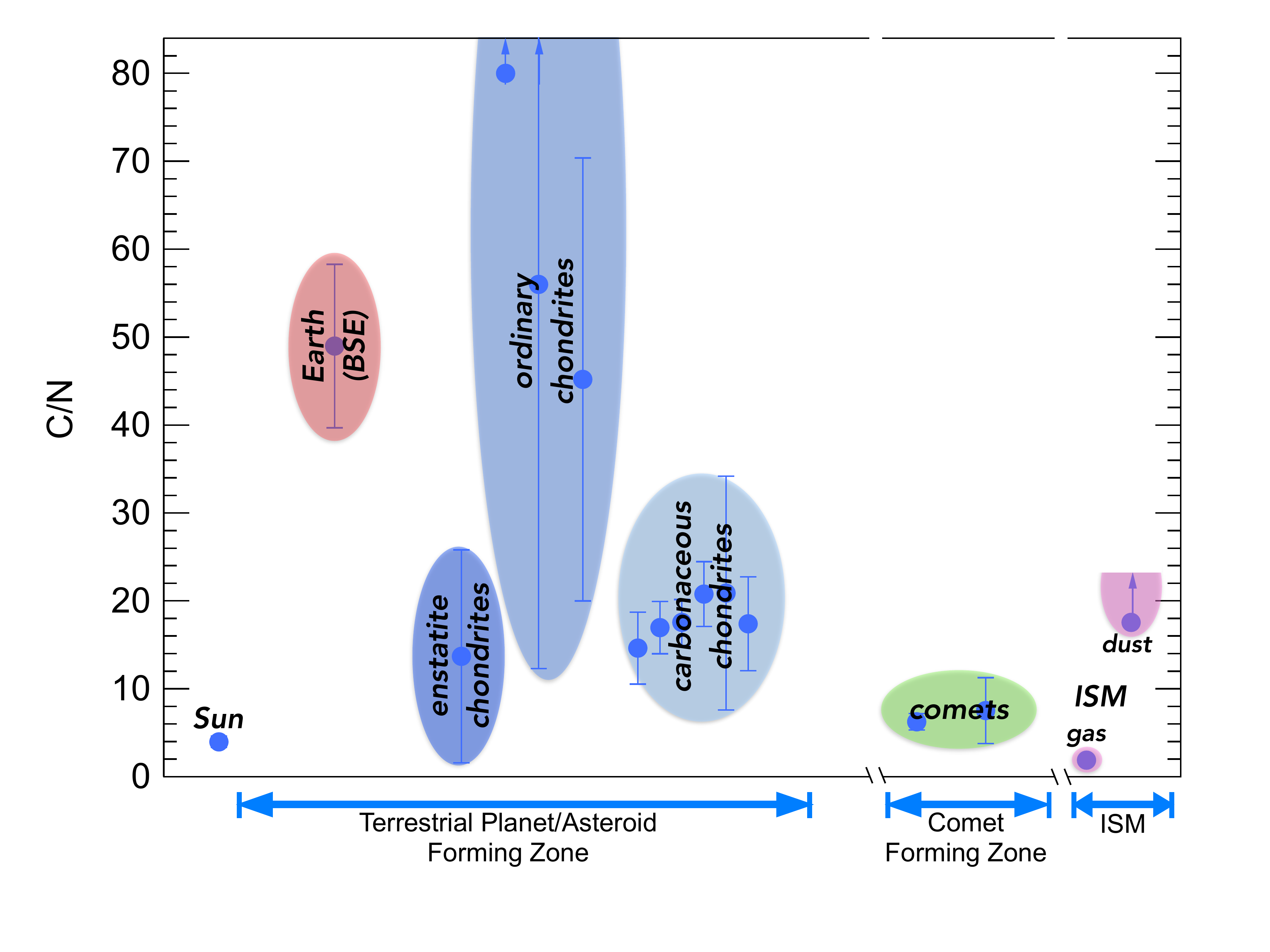}}
\caption{\small  Atomic ratios of bulk carbon relative to nitrogen in various solar system bodies and the ISM, based on the compilation in Table S1 and references in Supplementary Information.  The shaded ovals represent the range/errors in the measured data for each class.  Ovals that are not closed indicate that the listed value is an upper/lower limit.  For chondrites the errors represent the range of determined values, while for other bodies the errors are 1$\sigma$ measurement uncertainties.
For additional details an expanded version of this figure with labels for each point is provided as Fig.~S1.
\label{fig:CN}}
\end{center}
\end{figure}

%\begin{figure*}[ht]
%\centering
%\begin{subfigure}{.45\textwidth}
% \centering
%  \includegraphics[width=1.1\linewidth]{CoverSi-v2}
%  \caption{}
%  \label{c-si}
%\end{subfigure}
%\begin{subfigure}{.45\textwidth}
%  \centering
%  \includegraphics[width=1.1\linewidth]{NoverSi-v2}
%  \caption{}
%  \label{n-si}
%\end{subfigure}
%\caption{(a) Bulk elemental carbon to silicon atomic ratio and (b) nitrogen to silicon atomic ratio. References for calculations as given in Fig.~\ref{fig:CN}}
%\label{fig:siabun}
%\end{figure*}
\begin{figure}
\centering
  \includegraphics[width=0.9\linewidth]{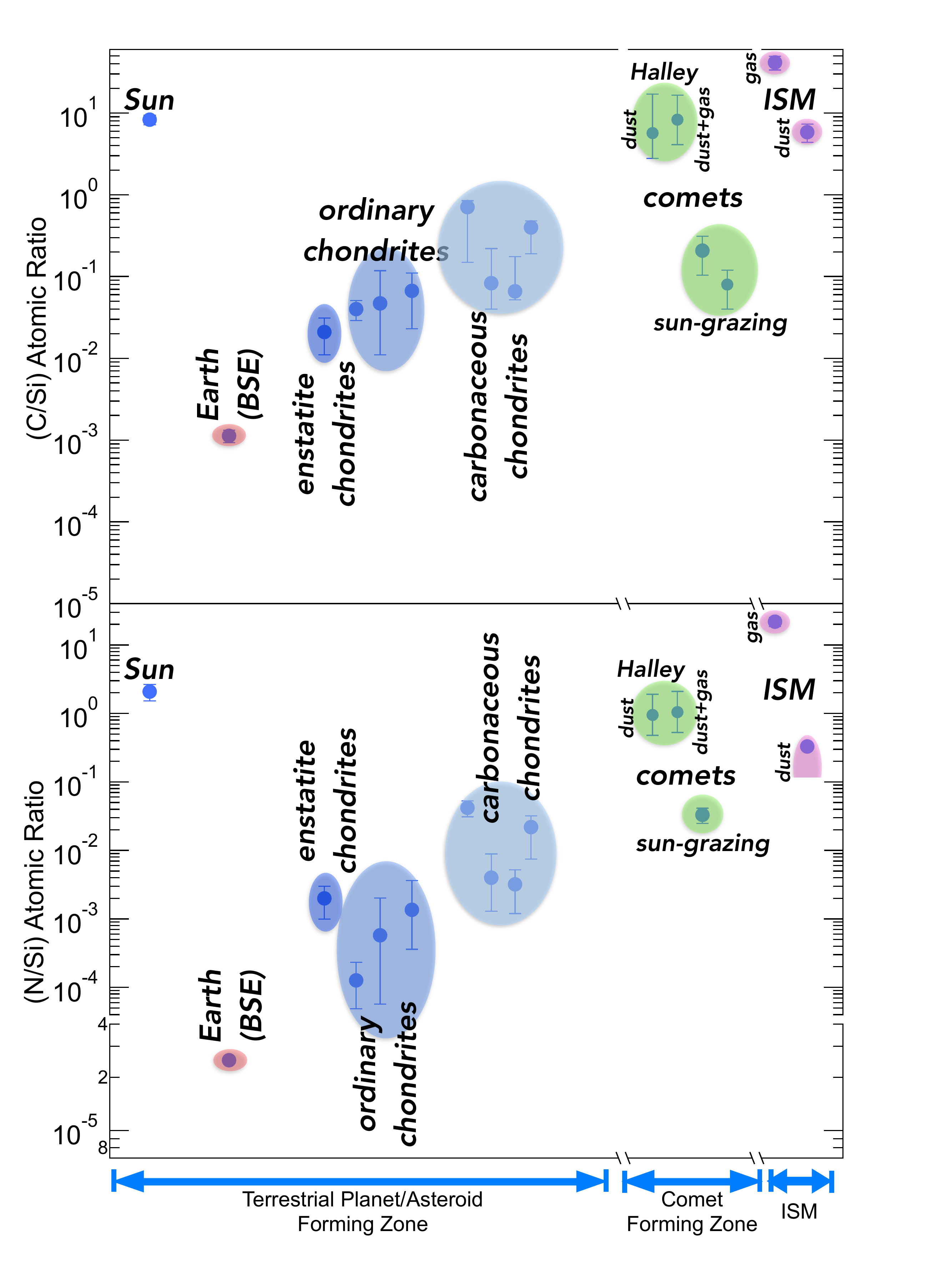}
    \vspace{-5pt}
\caption{\small (a) Bulk carbon-to-silicon  and (b) nitrogen-to-silicon ratios. Symbols are identical as in Fig.~1. 
References and an expanded figure with labels for each point in Supplementary Information.\label{fig:siabun}}
\end{figure}

These points tell a tale of an increasing C/N ratio from the cold
ISM, to cold cometary formation zones, to the inner nebula to planets.  While there is a broad overall trend, it is also clear that local processes can lead to variations within and between certain classes.
Whilst much of this secular variation must be due to high N
volatility, additional processes including preferential dissolution in
molten Fe,Ni-metal may contribute during assembly of planetesimals and planets. 
Notably the, BSE has the highest C/N ratios when compared to most solar
system material, except for ordinary chondrites. This high C/N ratio,
coupled to overall depletion of the BSE in absolute C and N
concentrations, indicate that processing and loss of these key
volatiles occurred during planetary assembly.  
Thus, there are several processes active early in the young solar system that control how planets are seeded with C and N. Below we
explore the processes responsible for these
trends, which must relate to the supply and fate of two of the key
ingredients of our biosphere.

\begin{figure*}
\begin{center}
\centerline{\includegraphics[angle=0, width=0.75\textwidth]{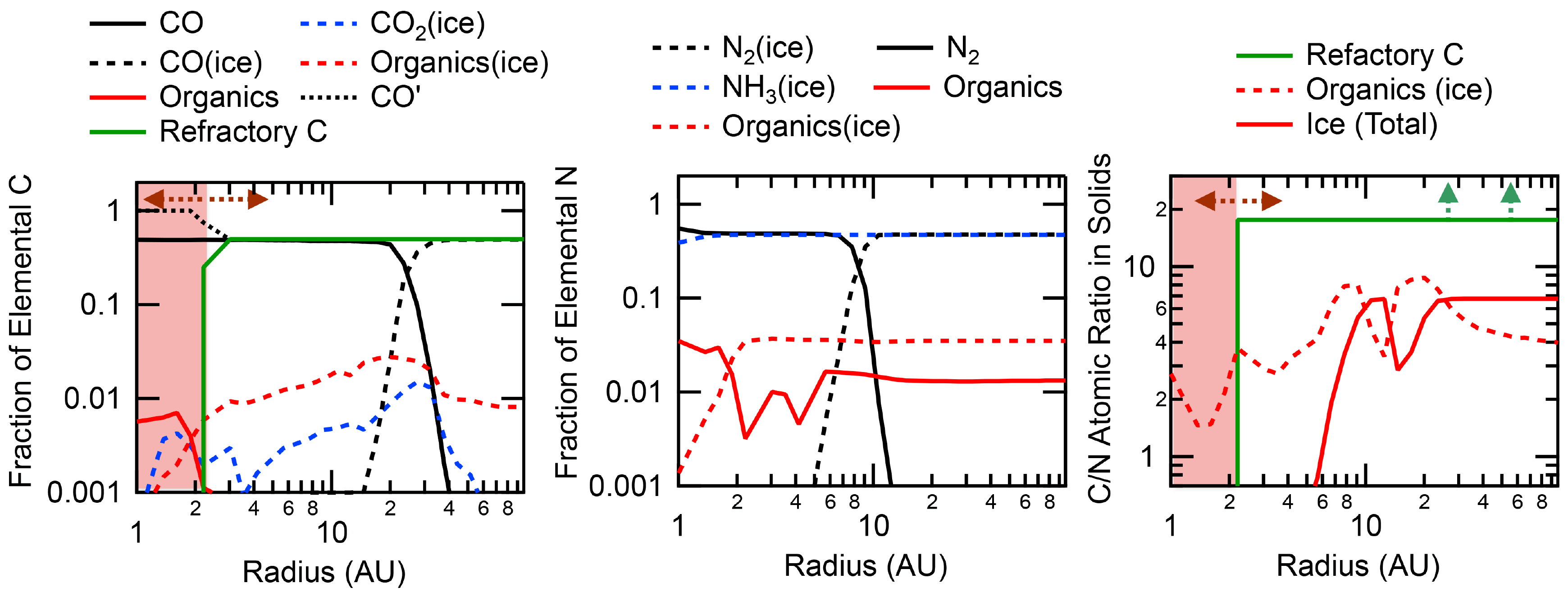}}
\vspace{-10pt}
\caption{\small Fraction of elemental C (left panel) and N (middle panel)  in molecular gases or ices from a kinetic chemical mode (age=1 Million years).  (right panel)  C/N ratio predicted for organic ices, the total ice mantle (organics and other volatiles shown in the other panels), and for refractory carbon (carbonaceous dust).    For C the evolution is complicated by the uncertain fate of carbonaceous dust which carries $\sim$50\% of elemental inventory.  Based on the composition of meteorites, some destruction of  carbon-rich dust is required.  The red shaded area represents the (uncertain) radii where carbon grains might be destroyed \cite{lbn10}.  Within this zone we artificially reduce C incorporated in carbonaceous dust to zero.  The principal product of the destruction is CO, which is artificially shown in the figure as CO$'$. \label{fig:kin} 
}
\end{center}
\end{figure*}

\section{C/N - ISM to Solar System Rocks and Ices}

\subsection{Elemental Condensation/Sublimation}

The depletion of carbon and nitrogen within meteorites and Sun-grazing comets is partially related to the high volatility of primary carriers within ices (e.g., CO, N$_2$), as opposed to the carbonaceous dust component.  In the case of C, it is further complicated by the fact that interstellar carbon grains provided to the disk need to be destroyed at some level.  This must be the case as the C/Si ratio of meteorites (and Sun-grazing comets) is an order of magnitude below that of the ISM dust (e.g., Fig.~\ref{fig:siabun}).  Carbonaceous dust destruction mechanisms likely include reactions with atomic O and OH (oxidation), photodissociation of small ($<$ 50 C,H atom) clusters \cite{gail02, lbn10, Montillaud14}, and parent body processing (see following section). Oxidation can destroy carbon grains, but not silicates, but requires high gas temperatures ($>$few hundred K). 
This process can be active on the surface of the inner disk and/or during the hot, early evolution stages leading to C depletion in solar system rocks \cite{gail02, lbn10}.  At greater distances from the Sun, or at later times, interstellar carbon grains need to be significantly preserved, or perhaps gain additional carbon by extraction from CO \cite{Bergin14}, to match the carbon content of comet Halley and IDPs.
%In the case of nitrogen, interstellar carbonaceous grains are a potential source of N for solar system solids.   
% This still leaves open the question regarding the overall nitrogen depletion of cometary ices.
The difference between comet Halley, IDPs, and the Sun-grazing comets may relate to the processes discussed above.\footnote{In the supplemental material we discuss the available information regarding the formation location(s) of Halley and the Sun-grazing comets.}

 To explore this chemical evolution, we present the main carriers of elemental C and N within a generic protoplanetary disk kinetic chemical model (see Supplementary Information) in Fig.~\ref{fig:kin}.     The temperature of microscopic solids from 0.5-3~AU will be uniformly above the CO condensation temperature of $\sim$20~K.  Nitrogen, primarily N$_2$, is also gaseous here, though NH$_3$ (condensation temperature of $\sim$80~K) has lower volatility than the main carbon ice carriers (CO, CO$_2$) and could provide nitrogen to some solids.
   % Thus, because 50\% of the carbon will not condense, there is no 50\% condensation temperature for carbon at these radii.   
%   A roughly similar story exists for nitrogen, with one difference, the presence of NH$_3$, which has lower volatility when compared to the main carriers of carbon in ices (e.g., CO, CO$_2$).
   In this example, it is only inside 2~AU that NH$_3$ ice begins to sublimate.   %This leads to the formation of N$_2$, which has comparable volatility to CO\cite{Bisschop06}.
Organics are present at these radii and, due to gas-phase chemical processing some of the C in CO and N in N$_2$ can be reprocessed in the gas leading to organic ices \cite{Bergin14}.  
%This is an important source for organic gas and vapor inside the CO snow line ($\sim$20 AU in this model) producing the few percent C and N in organics shown in Fig.~\ref{fig:kin}.  

  Thus, as seen in Fig.~\ref{fig:kin}, the carbon carriers with the lowest volatility and the primary carriers of C in inner solar system solids will be organic ices and/or carbonaceous dust grains.  These are the carriers of elevated C/N ratios into the warmer inner disk. We provide a brief discussion of the relative sublimation temperature of organic materials in the Supplementary Information.
%\footnote{The estimated sublimation temperature of organic ices depends on the binding strength to the surface; based on laboratory measurements\cite{Voumard95, Oberg09a}, this is on the order of 150-250~K for low molecular weight species (the latter corresponds to a binding energy of 7500~K, for a molecular mass of 60 amu, and a pressure of $\sim$10$^{-4}$ bar).  For carbon in organic ``refractory'' form we can use the literature on sublimation enthalpies for hydrocarbons which are $\sim 100$ kJ/mol\cite{Chickos02, Tabernero12}, with a corresponding sublimation temperature of $\sim 400$~K.   Some fraction of the C may be present as macromolecular, cross linked species, and would survive to even higher temperatures.}
 In the comet-forming zones (R $\gtrsim$ 5 AU), carbon is primarily found as carbonaceous dust and as CO ice ($>$20~AU) or CO vapor ($<$20~AU).  At least 50\% of the carbon resides in dust with the condensation/sublimation temperature $\ge$ 400 K.  For the case of nitrogen, in this model the inner 10 AU are dominated by NH$_3$ ice and N$_2$ gas, while beyond 10 AU there is even partitioning between N$_2$ ice and NH$_3$ ice.   

Based on this analysis, the C/N ratio in the starting materials for planets depends on the mixture of carbon in carbonaceous dust and organic ices coating dust grains.  Given the low carbon content of chondritic meteorites, the presence of interstellar carbonaceous dust grains might be time/distance variable in the disk.  In contrast, organic ices might be created {\em in situ} at all radii with C/N of 1-6 (Fig.~\ref{fig:kin}), while incorporating some carbonaceous dust would give higher values.   Comet Halley provides a limit for the carbonaceous dust and ice content in the ice giant planet-forming zone, with  C/N = 3.7--11.5.   Disk chemistry can thus implant C/N ratios ranging from $1$ to $\sim$12, values consistent with comets and the low end of the primitive chondrites.

\subsection{Loss During Parent Body Processing in the inner solar system}

A key issue in interpreting the C/N ratios of meteorites is that they are not pristine records of the dust and ice that accreted to form planetesimals in the solar nebula.  Meteorites may provide a good analog to such processes, as their parent bodies have been heated and differentiated, modifying both their C and N concentrations. \cite{huss06,brearley_messii}.   Chondritic meteorites, are believed to be direct aggregates of dust grains from the solar nebula, but were exposed to high temperatures (up to 1300 K) and experienced different degrees of thermal metamorphism, in some cases resulting in elevated C/N.  This does not mean that high C/N meteorites, such as ordinary chondrites (Fig. 1) were direct sources of high C/N materials delivered to Earth, as chemical and isotopic arguments rule out these samples as primary terrestrial precursors \cite{drake_righter}.    Rather, the processes that they experienced may have been analogous to processes that occurred on parent bodies of objects from which the Earth accreted.

Volatile species, including some organics, would decompose in meteorite parent bodies at high temperatures and the liberated C and N could migrate to the surface as a gas and be lost to space, or could be incorporated into metallic Fe,Ni (kamacite and taenite) owing to their siderophilic behavior \cite{hashizume98}.    Thus the abundances of both C and N are expected to diminish within bodies that were exposed to higher temperatures and more prolonged heating.  This is seen in both ordinary \cite{grady89,hashizume95,wieler06} and carbonaceous chondrites \cite{Pearson06};  additional evidence is found in the volatility of organic components and their carbon and nitrogen isotopic ratios \cite{Sephton03}.
% Sephton et al. \cite{Sephton03} found that lower C and N abundances correlates with low bulk $^{13}$C and $^{15}$N abundances.  As meteoritic organics are associated with $^{13}$C and $^{15}$N enrichments, this further supports the C/N abundances being determined by organic decomposition.  

The extent of C and N loss during parent body processing is determined by the initial compounds accreted, the chemistry that occurs during heating, and the intensity of heating.  For example, the vigor of C and N release from organics during thermal metamorphism depends on the redox conditions, particularly for C, which is incorporated into a gas under oxidizing conditions, but is stabilized in refractory compounds at more reducing conditions \cite{hashizume98}.  This could account for much of the variation in C/N ratio seen in ordinary chondrites.  However, enstatite chondrites have most of their N in the form of nitrides while carbon is found chiefly as graphite and carbides \cite{gw03}.  The reduced character of enstatite chondrites likely stabilizes solid C and N compounds to higher temperatures, inhibiting large-scale modification of the original C/N value.

Continued heating on parent bodies would lead to melting, and the formation of iron cores surrounded by silicate mantles and any C and N not lost to space would be distributed according to their low-pressure metal/silicate partition coefficients.  Growing planets accrete both silicate and metal portions of such a body, either in individual events or, if fragmentation occurs \cite{Chambers13}, in aggregate of multiple accretion events.  However, it is chiefly the volatile loss that occurs during heating that would determine the final inventory of materials delivered to a growing planet.  Thermally processed ordinary chondrites may thus show us the type of processing that occurred on the planetesimals and embryos accreted to the Earth: thermal evolution of small bodies derived from nebular dust and ice produced differentiated bodies with enhanced C/N ratios.

\vspace{-20pt}
\section{C/N - Forming Terrestrial Worlds}

Volatiles accreted to the Earth influence climate and biosphere development if they are available to the near-surface environment, either as remnants of the early primordial atmosphere or from material incorporated into the silicate Earth and later outgassed via volcanism.   Because the early atmosphere was susceptible to loss via impacts \cite{Genda05, Schlichting15}, the latter may be most important.  But, volatiles contained in silicate could be drawn down into the core during the era of silicate-metal segregation.   Consequently, the volatiles presently in the BSE have escaped loss to space or sequestration by the core, leaving them potentially available to the modern surface environment.

Current models of early planetary differentiation postulate that formation of the core and of an early massive atmosphere are associated with a deep magma ocean, with metal-silicate partitioning established at high mean pressures and temperatures \cite{Li01, Righter11} -- possibly at conditions that became more oxidizing with time \cite{Wood08, Corgne08}, and with a significant fraction of the planetary volatile budget released from the molten silicates into the atmosphere \cite{Kuramoto97, et08}.  Both C and N are moderately to highly soluble in molten Fe-rich alloy \cite{wood93, Dasgupta2013b, Roskosz13}  and comparatively insoluble in silicate \cite{Pan1991, Libourel03, Ardia13}. Thus, much of the Earth's initial C and N inventories likely segregated to its metallic core or were released to the atmosphere.   The quantitative effects of these processes on BSE C and N content, and on the C/N ratio, depends on partitioning between molten Fe-rich alloy and molten silicate, which in turn depends on the conditions of metal-silicate exchange and on the solubilities of C and N in silicate at the magma ocean-protoatmosphere interface.  Also, the effects of core segregation depend on the fraction of metal equilibrated with volatile-bearing silicate \cite{hirschmann12}, as much of the core may have segregated prior to bulk volatile delivery \cite{morby00}.

%, especially if the latter arrived principally during the end stages of mass accretion 
Experimental studies establish that C and N partition strongly into molten Fe-rich alloy coexisting with silicate melt.  Though values for both elements vary with experimental conditions and compositions, C partitions more strongly into metal than does N.  Characteristic mass partition coefficients for C ($D_{\rm C}^{met/sil}$) are 10$^{2}$-10$^{4}$  \cite{Eggler79, Dasgupta2013b, Chi14, Stanley14, Armstrong14}, compared to those for N ($D_{\rm N}^{met/sil}$) of $\sim$20 \cite{Roskosz13}. 
The  solubilities of C and N depend strongly on oxygen fugacity, with C more soluble than N in silicate melt under oxidizing conditions, and the opposite under highly reducing conditions \cite{Libourel03, Stanley14, Armstrong14}. 

\begin{figure}
\centering
  \includegraphics[width=0.99\linewidth]{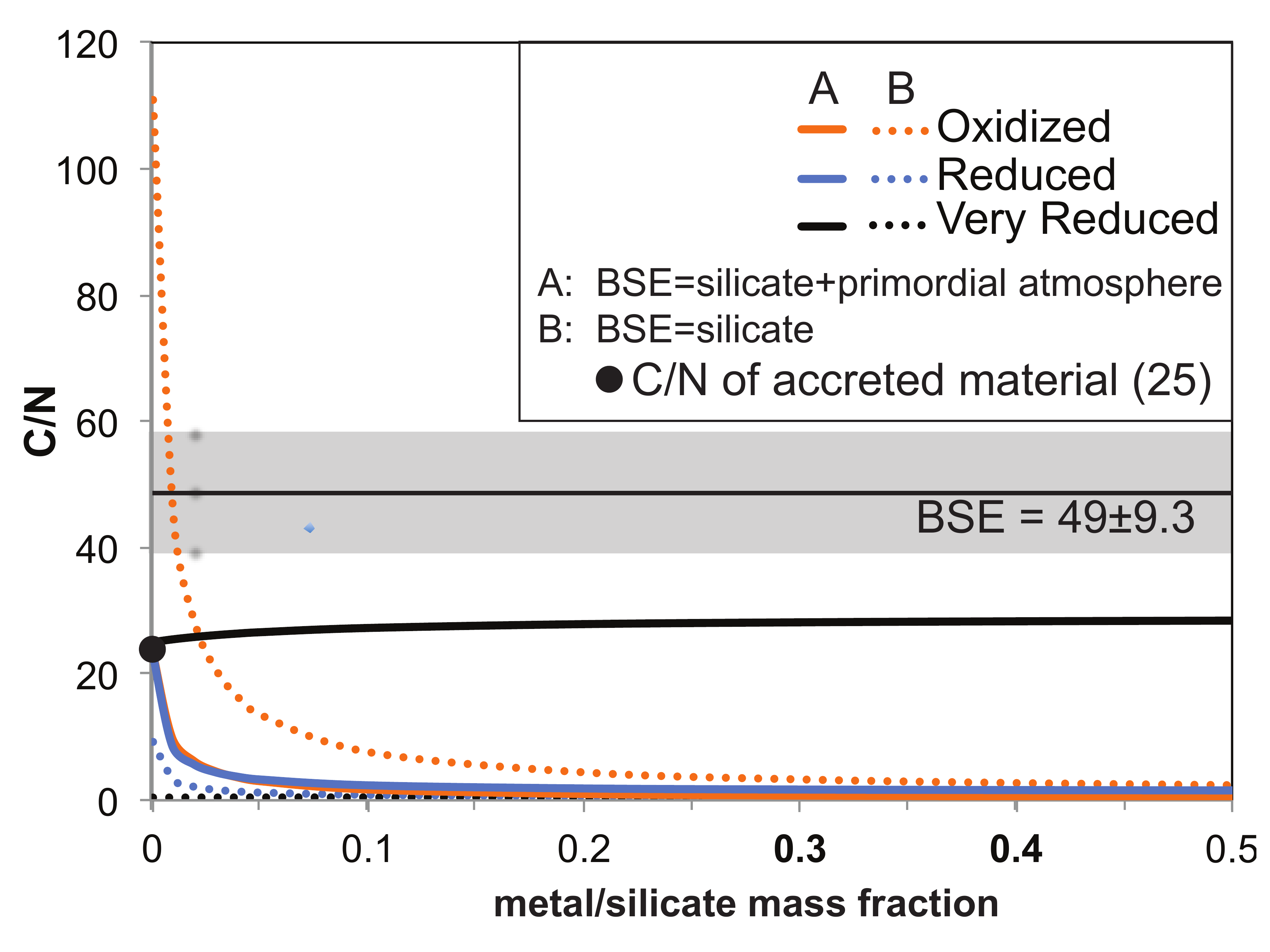}
  \vspace{-10pt}
\caption{\small Calculated C/N for the BSE resulting from a whole-mantle magma ocean equilibrated with an overlying atmosphere and variable amounts of core-forming metal \cite{hirschmann12}.  All calculations assume the proto-Earth begins with C/N $=$ 25 and these elements are initially present in the silicate.  (A) shows C/N for a BSE deriving from the sum of the magma ocean and the atmosphere, whilst (B) shows the BSE C/N derived only from the magma ocean, assuming that the primordial atmosphere is lost to space. ({\bf Oxidized}):  C in vapor is assumed to be chiefly CO$_2$ and N is dissolved as N2.  ({\bf Reduced}):   C in vapor is assumed to be chiefly CO \cite{Armstrong14}. N is dissolved as N$_2$.
 ({\bf Very Reduced}):  C is highly insoluble and N highly soluble. Quantitative solubilities for case 3 are poorly constrained, but for C and N must be very low and high, respectively.  See Supplementary Information for calculation details.
  \label{fig:cnearth}}
\end{figure}

Although terrestrial volatile accretion and differentiation were dynamic processes evolving over a range of conditions, insight can be gained from simple models that begin with a molten mantle in equilibrium with some fraction of core-forming metal and with an overlying atmosphere \cite{Kuramoto97, hirschmann12}.  
To track the evolution of the C/N ratio during planetary processing, in Fig.~\ref{fig:cnearth} we consider two end-member cases, both adopting an initial stage of chemical equilibrium among a fraction of core-forming iron-rich alloy or metal, a molten silicate mantle, and an overlying atmosphere, followed by the segregation and isolation of the core from the mantle. In case A, the atmosphere returns to the silicate mantle to form the BSE, whereas in the other, case B, the atmosphere is blown off to space and lost from the Earth.  Motivated by the earlier discussion, we assume an initial C/N ratio of 25.
%In all we consider three scenarios with different redox conditions. 

The most oxidizing condition  (orange curves, Oxidized) is a scenario in which carbon in vapor is assumed to be CO$_2$ and N in vapor as N$_2$.  We first discuss Case A where the magma-silicate Earth is enshrouded by a primordial atmosphere.   Here, the initial equilibrium leaves a large portion of N in the atmosphere with most C in the condensed phases and predominately in the core. Thus, if the primordial atmosphere was retained, the BSE will end up with a C/N ratio slightly reduced from the initial value, provided a small fraction of metal equilibrated with the mantle. The ratio approaches zero if the entire core (metal/silicate$\sim$0.5) equilibrated with the molten silicate and sequestered most C away.  In case B (atmosphere blow off) the BSE C/N ratio would be elevated with respect to the initial value if core sequestration is limited, but reduced if a large fraction of the core equilibrated with the mantle and strongly depleted the C of the molten silicate.

The second scenario (blue curves, Reduced) represents  moderately reducing conditions wherein carbon in vapor is assumed to be CO and nitrogen as N$_2$. If the atmosphere is retained, the results are qualitatively similar to the first case, but if the atmosphere is blown off this results in a BSE C/N ratio  that does not exceed the initial value even if no C was sequestered by the core. The lower than initial C/N ratio is derived from the smaller solubility of C in molten silicate. 

The final scenario (black curves, Very Reduced) represents a highly reducing condition where C is assumed to be highly insoluble, with a speculated solubility of 0.05 ppm/MPa, and N highly soluble, with a speculated value of 50 ppm/MPa. Contrary to the previous two cases, initial equilibrium among metal, silicate, and atmosphere leaves most N in the condensed phase and a significant portion in the silicate, whereas C resides primarily in the atmosphere. For case A, the BSE C/N ratio is slightly elevated from the initial value because some N was sequestered by the core and C escaped core sequestration by being part of the primordial atmosphere. On the other hand, for atmospheric blow off (case B), most C would be lost and the C/N ratio of the BSE would be markedly reduced.
%This would imply that C and N were accreted after the H-rich solar nebular had dissipated and after the main phase of core formation had completed. 

%We consider cases where the BSE forms from the silicate mantle plus the overlying atmosphere and instances where the latter is subsequently lost to space, leaving the BSE to be only the silicate portion of the magma ocean.  

 All models assume that $D_{\rm C}^{met/sil} \gg$  $D_{\rm N}^{met/sil}$ so the consequences for the BSE C/N are affected by: (a) the relative solubilities of C and N in silicate melt, and (b) the proportion of metal that equilibrates with the magma ocean.  As illustrated by Fig.~\ref{fig:cnearth}, if C is modestly soluble in silicate and N insoluble, as is true under oxidizing or modestly reducing conditions, then C is chiefly dissolved in condensed phases, and predominantly so in metal. The BSE (silicate + atmosphere) thus has greatly diminished C/N, even for small fractions of available molten alloy.  If, however, the amount of metal is small, then following atmospheric loss, the residual silicate mantle may have high C/N.  Alternatively, if C is highly insoluble and N very soluble, as may be true under very reduced conditions  \cite{Libourel03, Armstrong14}, then the proportion of N dissolved in condensed phases, including core-forming metal, is enhanced. The atmosphere is therefore carbon-rich, and so the BSE can have a high C/N ratio.  
 
 In sum, under most plausible scenarios, magma ocean-related core formation, combined with atmospheric blow off, produces much lower C/N ratios than observed.  The exceptions is a speculative circumstance of highly reduced conditions where N is  more soluble than C in silicate.  More likely,
% blow-off of a thick atmosphere developed over a nearly metal-free magma ocean.  
%The latter may be most consistent with scenarios that involve (a) late delivery of volatiles, multiple magma oceans \cite{Tucker14}, and atmospheric ablation from many smaller impacts following magma ocean solidification \cite{Schlichting15}.
N-rich atmospheric blow-off is necessary to elevate the C/N ratio from carbon depleted levels (Fig.~\ref{fig:cnearth}) to the current value.
 
% Moreover, planetary processes can increase the C/N ratio significantly only if metal/silicate/vapor equilibrium occurred under oxidizing conditions and involving a small fraction of core equilibrating with the mantle. 

\section{Discussion}

The C/N ratio provides diagnostic information regarding physical processes active during the earliest stage of planetary birth and assembly. The most likely initial carriers of C and N are low-to-moderate volatility organics and carbonaceous grains. The organics can derive in part from the dense ISM before stellar birth, but also can be created  inside the gas-rich  disk. Based on models and observations, disk chemistry and transport can imprint variable C/N ratios ($\sim$1-12) in ices comparable to those of comets and primitive meteorites. During the stages of planetesimal formation when bodies grow to large sizes ($>$ tens of km),  impact processing and thermal metamorphism can lead to greater disparity in the C/N content of pre-planetary material (Fig.~\ref{fig:CN}).  This results in greater C retention relative to N, leading to great variability in the C/N ratio between planetesimals of different sizes and within individual bodies.  Such effects are well-illustrated by the large dispersion in the C/N ratios measured in ordinary chondrites.  Although cosmochemical constraints indicate that ordinary chondrites are not the sole source of Earth's volatiles \cite{drake_righter}, analogous processes may have elevated C/N ratios on many planetesimals.  As a consequence, the mean C/N ratio of materials accreted to the Earth may have been greater than those evident in primitive bodies such as carbonaceous chondrites or comets. 

Even assuming that the mean C/N ratio of material accreting to the growing Earth was enhanced, explaining the high C/N of the BSE remains a challenge because core formation should drastically reduce C/N (Fig~\ref{fig:cnearth}).  Thus, substantial loss of N to space seems to be required.  This may be most consistent with scenarios that include: (a) late delivery of volatiles, chiefly from comparatively oxidized and metal-poor bodies \cite{Rubie15}, thereby adding volatile-rich material to the mantle without loss of metal to the core, (b) multiple magma oceans punctuated by large atmospheric loss events \cite{Tucker14}, and (c) atmospheric ablation from many smaller impacts following magma ocean solidification \cite{Schlichting15}.  The high C/N ratio of the BSE therefore appears to be a sensitive indicator of the balance of volatile accretion and loss during the final stages of the Earth's assembly.  Viewed more broadly, such a scenario will likely result in a highly variable supply and retention of these key ingredients to the surface reservoirs of terrestrial worlds.

\begin{acknowledgments}
This work was supported by funding from the National Science Foundation grant
AST-1344133 (INSPIRE) to EB, GB, JL, and MH; FC was supported by NASA grant NNX12AD59G.  We are also grateful to I. Cleeves for providing
the kinetic chemical calculations from published work and to both anonymous referees for providing useful feedback that improved this manuscript.
\end{acknowledgments}

\newpage
\section*{Supporting Information}

\section{Bulk C and N Abundance Compilation and Estimation}

Below we summarize the information compiled and estimated for
Figures~1 and 2 in the main manuscript.  For completeness we provide
expanded versions of these figures here in 
Fig.~\ref{fig:s1} and Fig~\ref{fig:s2}.  Table~S1 provides our compilation
in tabular form.

\subsection{Terrestrial Worlds}\hspace{6pt}The Bulk Silicate Earth (BSE) assumes, for simplicity, that the Earth is comprised of two components: the core and the silicate mantle.  In this form the BSE is the entire Earth, including its fluid envelopes (oceans and atmosphere) minus the core.  Previous estimates of the C/N of the BSE differ markedly.  Marty\cite{Marty12} estimated a C/N molar ratio of 365$\pm$282, whilst in contrast Halliday's preferred model \cite{Halliday2013} amounts to 46$\pm$17.   This factor of 8 difference led us to revisit constraints on BSE C/N, using an approach somewhat different from previous work.

The BSE C and N inventories  derive from a combination of the surface and mantle reservoirs (Table S1). For the surface reservoir, we take total carbon to be 9.3 $\pm$0.9 $\times 10^{22}$ gm \cite{dh10} and total nitrogen to be 6.4$\pm 1.1 \times 10^{21}$ gm \cite{Goldblatt09}, yielding a C/N molar ratio of 16.9$\pm$0.3. For the interior reservoir we consider two separate sources, based on the well-established assumption that the mantle consists of a gas-depleted source (DM) that produces mid-ocean ridge basalts and a gas-enriched source (EM) related to oceanic island basalts \cite{Allegre96}.

We assume that the depleted reservoir constitutes between 30-80\% of the mantle mass \cite{Workman05}, and estimate the C concentrations of each reservoir based on C/Nb and C/Ba ratios of oceanic basalts compiled by Rosenthal et al. (in press), yielding 20$\pm$7 and 165$\pm$55 ppm for the DM and EM, respectively.  

Marty \cite{Marty12} and Halliday \cite{Halliday2013} estimated mantle N concentrations in part from the estimate of mantle C/N of 535$\pm$224  Marty and Zimmerman \cite{Marty99}.  However, this ratio derives from gas liberated from basalt crushing experiments.  Because N$_2$ fractionates from CO$_2$ substantially during vesiculation, C/N ratios of basalts and their vapor bubbles vary by more than two orders of magnitude \cite{Marty99, Cartigny01}, complicating reconstruction of source C/N.  Therefore, we estimate the N concentrations of each reservoir based on N/Ar ratios.  Unlike C/N, N/Ar ratios of vesicles in basalt do not change significantly owing to degassing \cite{Marty99, Cartigny01}. 
The mass of Ar in the mantle is constrained by the BSE inventory of K, 280$\pm$60 ppm, which over 4.55 Ga has produced 1.55$\pm 0.33 \times 10^{20}$ gm of $^{40}$Ar  \cite{Arevalo09}.  Subtracting atmospheric \cite{Turekian59} and crustal \cite{Lassiter04} Ar masses of 6.6 and 0.9 $\times 10^{19}$ gm, respectively, leaves 8.00 $\pm 3.3 \times 10^{19}$ gm of Ar in the present-day mantle \cite{Arevalo09}.  Because the N/Ar ratios of the depleted (DM) and enriched mantle (EM) reservoirs are seemingly distinct (N/Ar $= 74\pm56$ in DM and 105$\pm$35 in EM; \cite{Marty03}), conversion of the bulk mantle Ar mass to the bulk mantle N requires estimates Ar concentrations in each reservoir.  Taking the DM to have 3.3$\pm$1.6 ppb Ar (Marty and Dauphas \cite{Marty03}, but also consistent with Allegre et al. \cite{Allegre96} and Ballentine \cite{Ballentine2002}) allows calculation of the concentration of Ar in EM by subtraction from the bulk mantle Ar reservoir. 
The above information is sufficient to calculate the C/N ratios of the principle terrestrial reservoirs (surface reservoir, bulk mantle, BSE) (Table S1), taking in to account the uncertainties in individual concentrations and ratios must be  that one accounts for the uncertainties from each constraint.  This is done with a Monte Carlo simulation, as illustrated in Fig.~\ref{fig:s3}, which yields C/N ratios for the bulk mantle and BSE of 80  and 46$\pm$9, respectively. The mantle ratio is reported as a lognormal distribution because the arithmetic mean and standard deviation are strongly affected by simulations that produce very small mantle N (and thus extreme C/N ratios). This BSE C/N estimate is in nearly exact agreement with the previous estimate by Halliday \cite{Halliday2013}, 46$\pm$17, and markedly lower than that by Marty \cite{Marty12}.

Elemental abundances for the atmospheres of Venus and Mars are taken from the compilation of Hailliday et al. \cite{Halliday2013}.  
%Bulk silicate Venus is 23 wt.\% \cite{Fegley03} and Mars 21 wt.\% \cite{Halliday2013}.   
 The atmospheres of Mars and Venus
have C/N comparable to the Earth's surface (Supplementary Fig.~S1), though inferences for these
planets is limited by lack of constraints on interior reservoirs.   In the small sample of Martian meteorites the C/N ratio appears to vary, but are elevated above the solar value \cite{gw03}.

\subsection{Meteorites}\hspace{6pt}
Chondritic meteorites are a class of meteorites that escaped differentiation. Their bulk compositions generally match that of the solar photosphere, except for the most volatile elements.  These meteorites are broken into groups based on their bulk chemical, mineralogical, and oxygen isotopic compositions. In addition, the proportion and sizes of the different nebular solids they contain can be used to distinguish among the groups.  These groups are also categorized into three distinct classes of chondrites: the Enstatite, Ordinary, and Carbonaceous chondrites (EC, OC, and CC, respectively).  While chondrites are thought to be aggregates of solar nebula dust, these meteorites have experienced different degrees of planetary processing in the form of thermal metamorphism and aqueous alteration.  Aqueous alteration was most pervasive in the CI and CM chondrites, resulting from the melting of ice and reaction of rock with liquid water at temperatures $<$400-500 K.  While evidence for the action of water in other chondrites has been reported, thermal metamorphism, with meteorites reaching temperatures of up to 1300 K, is much more common in the EC, OC, and some of the other CC (notably CV and CO) meteorites.
 
For meteoritic C/N ratios we have used the values determined by Alexander et al. \cite{Alexander13, Alexander14} for CI, CM, CR, and CO chondrites,
while the data for CV chondrites are taken from the work of Pearson et al. \cite{Pearson06}.  For these data points we have determined the weighted mean and errors based upon the individual meteoritic measurements.  Thus the error represents the range of the data relative to the errors of the individual measurements.   For carbon and nitrogen elemental ratios relative to silicon we have used the bulk abundance measurements of Wasson \& Kallemeyn \cite{wk88}, with an assumed uncertainty of 20\% which is based upon their discussion of the relative uncertainties within carbonaceous chondrite groups.  We also extended the error bar range based on the variation seen in the C and N content within a given class from the references above.    For ordinary chondrites, information was gathered from different sources often studying the same meteorite but in different elements \cite{kc78, grady89, Sugiura92, Sugiura98, hashizume95}.  The error reflects the range of estimated C/N ratios for each class.    Abundances relative to silicon were estimated using the range of C and N from the previous sources for a given class and the average Si abundance for that class taken from Lodders \& Fegley\cite{Lodders98}.  For both ordinary and carbonaceous chondrites the error bars thus represent the variation of C and N rather than that of Si.

%GAB- Gaseous state refers, of course, to the coma, not the nucleus at larger heliocentric distances, so I tried to clean this up a bit.
\subsection{Comets}\hspace{6pt}Abundances for comet Halley rely on compilations/analysis of Fomenkova \cite{Fomenkova99}, Jessberger \cite{Jessberger99}, Delsemme \cite{Delsemme91}, and Wyckoff et al. \cite{wte91}.   In situ analysis of the refractory carbonaceous component was performed by instruments on board Vega-1, Vega-2, and Giotto spacecraft that performed fly-by's within the coma and that, with further analysis, provided elemental abundances.   For carbon, nitrogen, and silicon, Fomenkova \cite{Fomenkova99} state that gas and dust abundances of Halley for carbon are solar, nitrogen is a factor of two below solar, while silicon is within a factor of 2 of solar. Jessberger\cite{Jessberger99} states that the rock forming elements are within a factor of two of solar and chondritic; thus we adopt a factor of 2 error on the Si abundance which then dominates that calculation. For carbon we assume that 36\% is bound in molecular volatiles \cite{Delsemme91}, while only 10\% of the nitrogen inventory is tied up in such species \cite{wte91}.  Assuming solar abundances the C/N (gas and dust) ratio is (C/N)$_{\odot}$/2 and we adopt a 50\% error.  The evaporation of Sun-grazing comets provide another unique estimate of bulk cometary abundances.  Two comets have measurements of C and Si (C-2003K7 and C/2011 W3:Lovejoy) \cite{Ciaravella10, McCauley13}, while Comet Lovejoy has an additional measurement of nitrogen (provided by J. Raymond and given here).   Errors are assigned a value of 50\% which is based upon the estimated range in the C/Si ratio for C-2003K7\cite{Ciaravella10} and we extend this to the other measurements.  

Regarding the formation zone of Halley and the Sun-grazing comets, we note that the two Sun-grazing comets could be fragments of the same progenitor \cite{Sekanina12}.  However it is likely that the larger parent(s) originated within the Oort cloud \cite{Bailey92}, which itself is populated from comets formed near the giant planets and perturbed to larger orbits \cite{Duncan87}.  The origin of Halley-type comets has been a matter of debate; however the recent detection of a trans-Neptunian object in a highly inclined retrograde orbit  suggests a source region for Halley-type comets  with  a primordial origin at large distances ($>$ 50 AU) \cite{Gladman09}.

%GAB- When we state "at high extinction" I presume we mean A_V > 1, not less than? Changed here.
\subsection{Interstellar Medium}\hspace{6pt}In the ISM, elemental abundance measurements can be obtained via spectroscopic techniques towards low density ($<$1000 cm$^{-3}$) atomic gas in front of bright stars. These gas-phase data are compared to some reference abundance measurements, either the Sun or young stars (which likely represent the original composition of gas in their galactic location).  For a given element the deficit in abundance from the gas when compared to the solar/stellar standard is assumed to be sequestered into dust grains.  The presence of dust along these lines of sight is confirmed by its differential absorption effects on the broad band stellar spectrum.
 
 Thus, Jensen et al. \cite{Jensen07} surveyed 30 sight lines in transitions of atomic nitrogen and compiled abundances from previous work.
They find a gas phase nitrogen abundance at high extinction ($A_V > 1^m$) of 49 $\pm$ 4 ppm (relative to H) and, using adopting B stars as the standard, find that 14 ppm is missing from the gas phase.  This missing nitrogen could potentially be incorporated into the carbonaceous dust component and represents our limit for nitrogen in the solid state.

Based on elemental abundances measured toward sight lines that intersect diffuse interstellar gas/dust, the wavelength dependent extinction of starlight, and spectral features with unidentified carriers (e.g. the 2200 \AA\ extinction bump and mid-IR emission bands), it is a thought that significant percentage of interstellar carbon resides in some refractory component.  In 5 sight lines with high average density (log$_{10}\;\langle{\rm n_H/cm^{-3}}\rangle > 0.5)$, Parvathi et al. \cite{Parvathi12} estimate that 371 $\pm$ 37 ppm (relative to H) of carbon resides in interstellar dust grains, assuming B star standard abundances.  Along the same, higher density sight lines the abundance of gas-phase carbon is 94 $\pm$ 13 ppm.   The recent dust models of Jones et al. \cite{Jones13} and Chiar et al.\cite{Chiar13} adopt 233 ppm and 150 ppm, respectively.  For our purposes it is clear that there is a large amount of carbon in interstellar dust grains and we adopt 250 $\pm$ 50 ppm to account for the disparate abundance estimates.

As noted in the main text the median value of C/N ratio of interstellar ices is $\sim$ 12 \cite{oberg11_c2d}.  These measurements do not account for N or N$_2$,
 which are infrared inactive. Based on the similar sublimation temperature between CO and N$_2$ \cite{Bisschop06}, we  assume (CO/N$_2$)$_{ice} \sim 1$.  Further, assuming all N is in N$_2$ and 50\% of C in CO,  gives a C/N ratio for ices of $\sim$1.8.
 
 \section{Assumptions for Equilibrium Calculations}

 In our calculations in Fig.~\ref{fig:cnearth} we made the following assumptions. {\bf Oxidized}:  $D_{\rm C}^{met/sil} =$ 500, $D_{\rm N}^{met/sil} =$ 20, C in vapor is assumed to be chiefly CO$_2$, with depth-averaged C solubility in silicate liquid  of 1.6 ppm/MPa \cite{Pan1991},  N is dissolved as N2, with depth-averaged solubility=1 ppm/MPa \cite{Libourel03}. {\bf Reduced}:  $D_{\rm C}^{met/sil} =$ 1000, $D_{\rm N}^{met/sil} =$ 20, C in vapor is assumed to be chiefly CO, with solubility equal to 1 ppm CO/MPa \cite{Armstrong14}. N is dissolved as N$_2$, with solubility=1 ppm/MPa \cite{Libourel03}.
 {\bf Very Reduced}  $D_{\rm C}^{met/sil} =$ 1000, $D_{\rm N}^{met/sil} =$ 20, C is highly insoluble (0.05 ppm/MPa) and N highly soluble (50 ppm/MPa). We note that quantitative solubilities for case 3 in the figure lack constraints;  for C and N they must be very low and high, respectively.

\section{Kinetic Chemical Model}

Lodders\cite{lodders03} explored the chemistry in the solar system within the framework of the 50\% condensation temperature for thermodynamic equilibrium. 
% For carbon this depends on the local conditions with CO favored for $T > 650$~K and CH$_4$ at lower temperatures\cite{lp80, lodders03}; both of these species have low condensation temperatures and CH$_4$ is the relevant carrier with a condensation temperature of 41~K.   Nitrogen has a similar disposition between N$_2$ and NH$_3$ with ammonia favored at lower temperatures ($\sim 300-400$~K)\cite{lp80}.  Thus its condensation within the equilibrium framework would be set by NH$_3$ with a value estimated to be 131~K\cite{lodders03}. Clathrates have also been suggested as potential carriers, particularly for cometary nitrogen\cite{Iro03, Mousis12}
%However, clathrates have very specific formation conditions which may not be globally available\cite{ls85}.
%This concept implies path independent results given conditions.  
This concept implies that the chemical species produced in the disk are independent of the starting materials. 
However,as discussed by Lodders, the history and carriers matter for reactive volatile elements.  Thus, the 50\% condensation temperature is not a useful guide in this case, as thermodynamic equilibrium will not be achieved and we adopt a kinetic model to explore the evolution of these elements.

The kinetic chemical model presented in Fig.~3 of the main text is a compilation of chemical
abundance predictions from the model of Cleeves et al. \cite{cleeves13a}.   The physical structure is a 2D azimuthally symmetric
model that is motivated by the observations of solar analog protoplanetary disks.    Greater details regarding
the exact physical prescription can be found in the references listed above.  For completeness the basic physical structure is provided below in Fig.~\ref{fig:s4}.  Our methodology is to take an adopted dust density structure constrained by observations.  The gas density in this model, when compared to the dust, has a vertically integrated gas to dust mass ratio of 100.  However, in practice the gas to dust mass ratio is higher in the upper layers and reduced in the midplane to mimic the effects of dust settling.   With the gas and dust density prescribed we determine the dust temperature structure using the radiative equilibrium code TORUS \cite{Harries00}.  For this calculation the central star is assumed to have mass of M = 1.06 M$_{\odot}$, R = 1.83 R$_{\odot}$, and an effective temperature of 4300 K.  Due to high densities and frequent dust-gas collisions, the gas temperature is coupled to the dust for much of the disk mass, but is elevated at the surface due to stellar X-ray heating.  Within the framework provided by the dust and gas density we determine the UV and X-ray photon fields.  For the input spectra in the UV we use the spectrum measured towards TW Hya, the closest star/disk system.  The X-ray spectrum is estimated  assuming a two-temperature optically thin plasma with one component at T $= 9$~MK and another at 30 MK, and propagated assuming X-ray photoabsorption cross-sections.  
The resulting distribution of UV and X-ray flux is then calculated using the Monte-Carlo model developed by Bethell \& Bergin \cite{bb11a}.
For more explicit details and complete references the reader is referred to Cleeves et al.\cite{cleeves13a}.

The chemical model includes gas-phase processes, gas-grain interactions, and a limited set of grain surface reactions for the main carriers of C, O, and N linked through a network of $\sim$5900 reactions assuming initial chemical abundances \cite{cleeves13a}.   To create Fig.~3 we ran the model for a million years and compiled the predicted abundances.   Based on these predictions we summed up the abundances of all carriers of carbon and nitrogen to isolate the main carriers.  These are (both gas and ice phase): CO, CO$_2$, N$_2$,  NH$_3$, and organics.   Refractory carbon contains $\sim$50\% of elemental carbon in the interstellar medium and we assume it is oxidized \cite{lbn10} interior to $\sim$2 AU where it is released as CO (listed as CO$'$ in Fig.~3).  The location in the disk where this process is active is highly uncertain.  The main organic carriers of C, where the abundances were summed for the figure (separately as gas and ice) are: HC$_3$N, C$_3$, HCN, CH$_3$OH, C$_3$H, HNC, C$_4$H$_2$, C$_2$H$_4$, OCN, C$_3$O, H$_2$CO, and CH$_4$;
for nitrogen the organic carriers are: HC$_3$N, HCN, HNC, and H$_2$CN.  An important issue is the presence of NH$_3$ in the ices even at 2 AU.  This is due to the relatively high sublimation temperature of NH$_3$ compared to N or N$_2$ (at relevant pressures, NH$_3$: $\sim 80$~K for pure ice or higher if bound with water; N$_2$: $\sim 20$~K)   \cite{Bisschop06, md14}.
 In this light the depletion of nitrogen in cometary ices is difficult to interpret.  However, a recent exploration of nitrogen in protoplanetary disks can match the nitrogen content in cometary ices provided nitrogen is initially supplied to the disk as atomic N, which is then converted to gaseous N$_2$ \cite{Schwarz14}.

The estimated sublimation temperature of organic ices depends on the binding strength to the surface; based on laboratory measurements\cite{Oberg09a}, this is on the order of 150-250~K for low molecular weight species (the latter corresponds to a binding energy of 7500~K, for a molecular mass of 60 amu, and a pressure of $\sim$10$^{-4}$ bar).  For carbon in organic ``refractory'' form we can use the literature on sublimation enthalpies for hydrocarbons which are $\sim 100$ kJ/mol\cite{Tabernero12}, with a corresponding sublimation temperature of $\sim 400$~K.   Some fraction of the C may be present as macromolecular, cross linked species, and would survive to even higher temperatures.

\end{article}

\newcommand{\beginsupplement}{%
        \setcounter{table}{0}
        \renewcommand{\thetable}{S\arabic{table}}%
        \setcounter{figure}{0}
        \renewcommand{\thefigure}{S\arabic{figure}}%
     }
\beginsupplement

\begin{figure}
\includegraphics[angle=0, width=.9\textwidth]{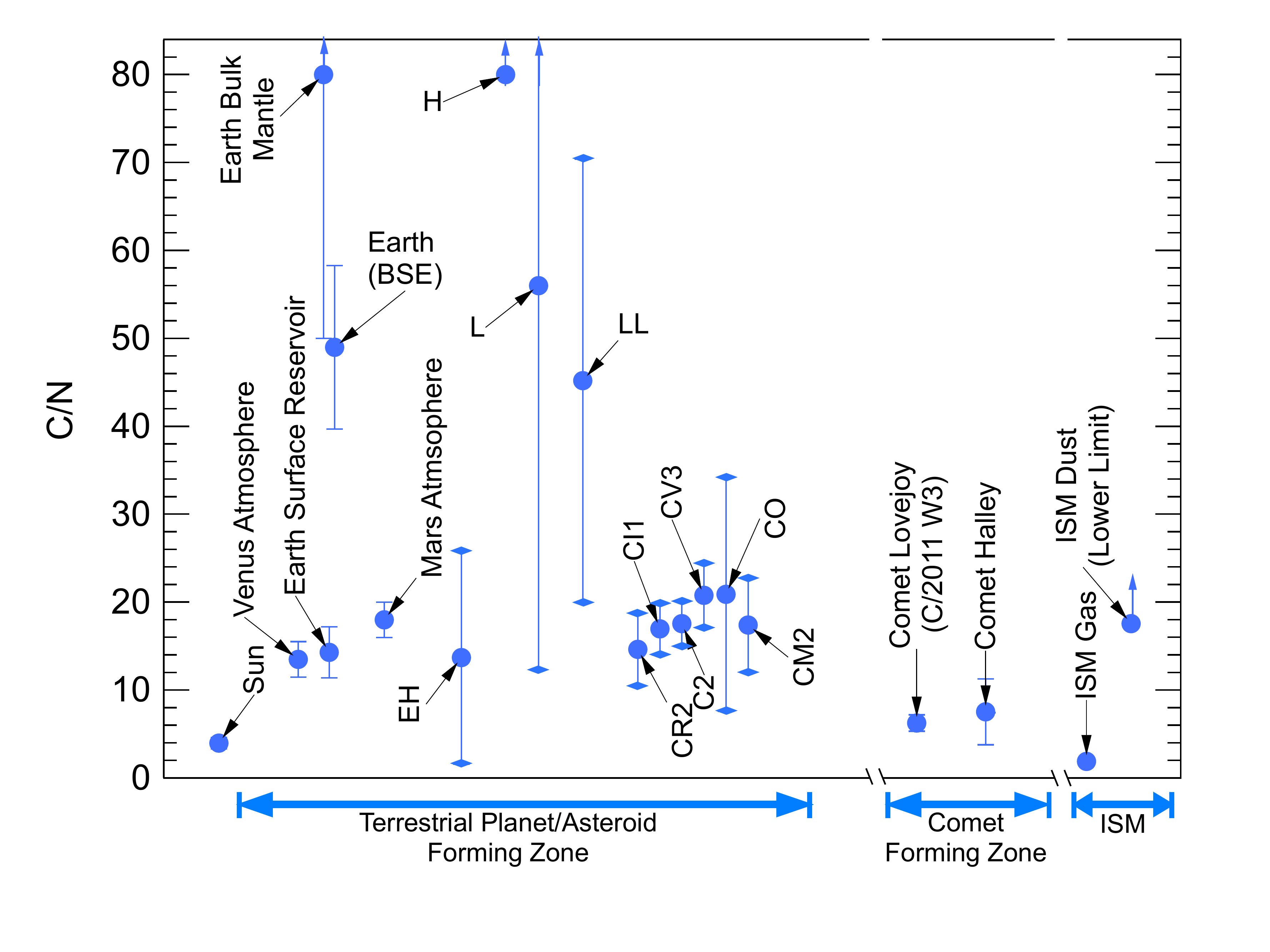}
%\vspace{-20pt}
\caption{\small Expanded version of Fig. 1 in the main manuscript.  Ratio of bulk (or as noted) carbon relative to nitrogen in various solar system
bodies and ISM. Errors denoted with a diamond represent ranges within the samples and not the measurement error.  Normal errorbars represent the 1$\sigma$ uncertainty in the measurement.  References for this compilation are given here.\label{fig:s1}}
\end{figure}

\begin{figure}
  \includegraphics[width=0.9\linewidth]{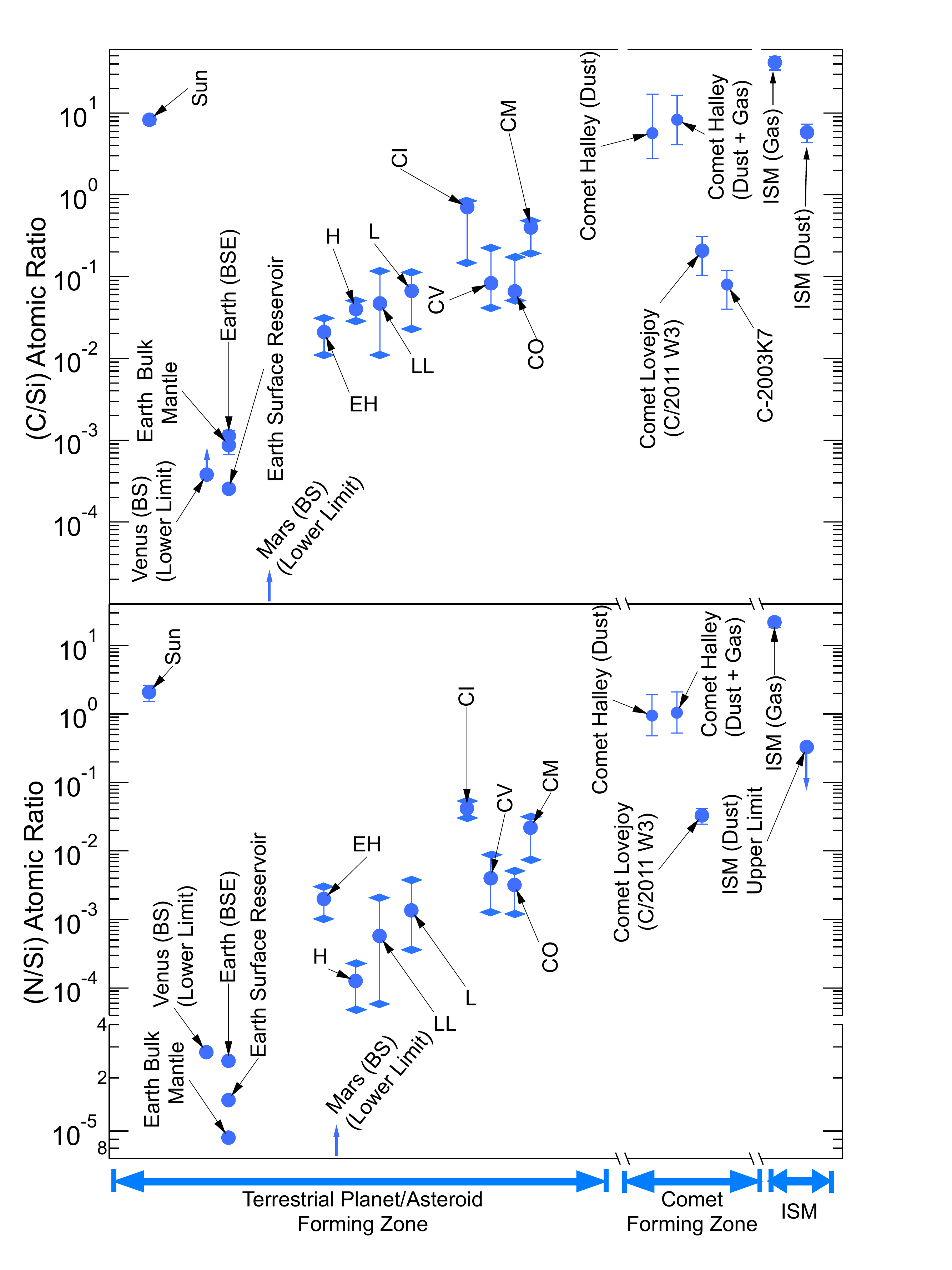}
\caption{\small Expanded version of Fig. 2 in the main manuscript.   (a) Bulk elemental carbon to silicon atomic ratio and (b) nitrogen to silicon atomic ratio. Errors denoted with a diamond represent ranges within the samples and not the measurement error.  Normal errorbars represent the 1$\sigma$ uncertainty in the measurement.  References for this compilation are given here.\label{fig:s2}}
\end{figure}

\begin{figure*}[ht]
\begin{center}
\centerline{\includegraphics[angle=0, width=.9\textwidth]{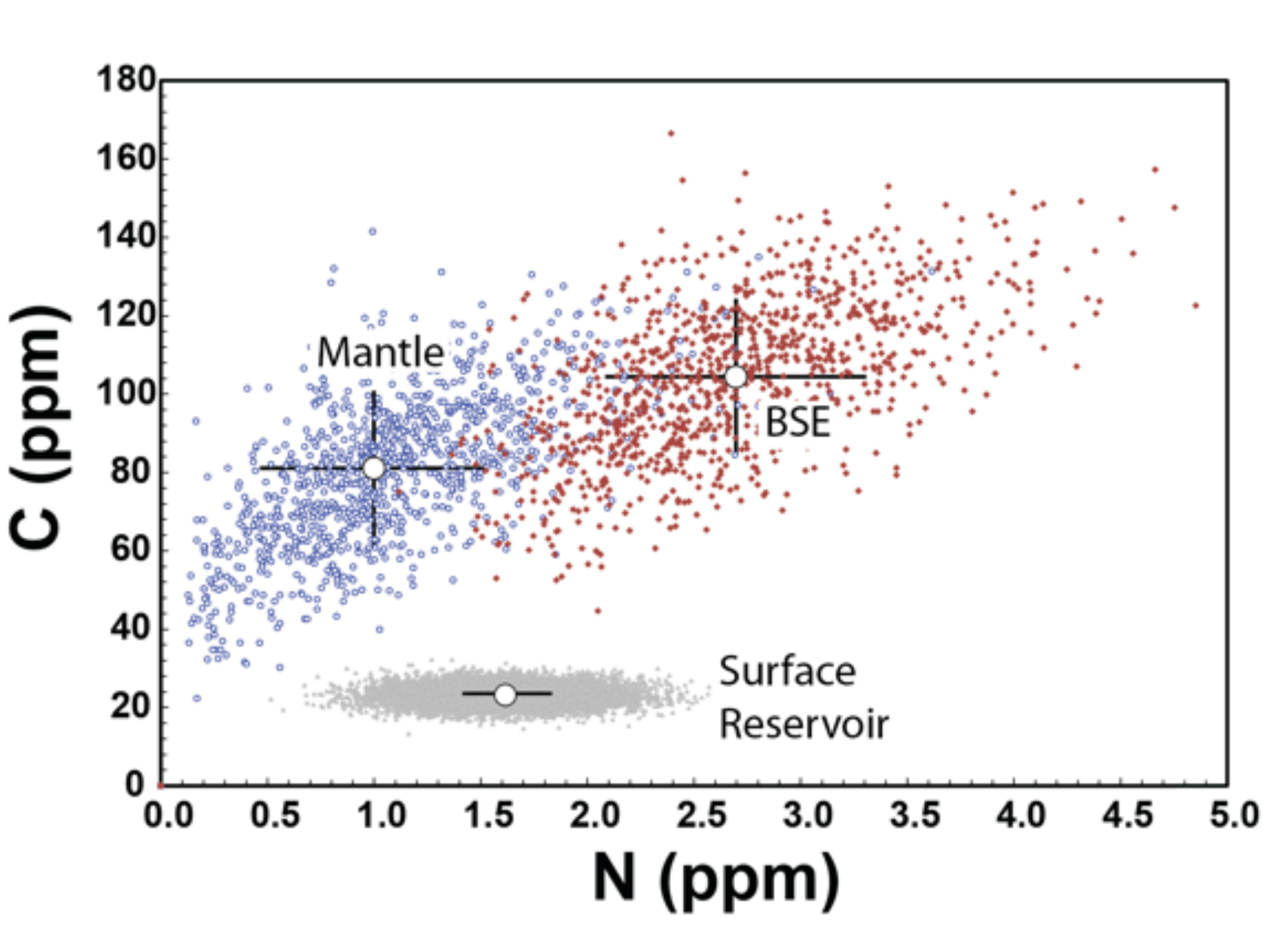}}
\caption{Results of Monte Carlo simulation showing compositions of surface reservoir, bulk mantle, and bulk silicate Earth (BSE).  Concentrations are in ppm by weight, calculated from the mass of C and N in each reservoir divided by the mass of the BSE (4 $\times 10^{27}$ gm).  The Monte Carlo simulation was performed with 1000 individual calculations. \label{fig:s3}}
\end{center}
\end{figure*}

\begin{figure*}[ht]
\begin{center}
\centerline{\includegraphics[angle=0, width=.9\textwidth]{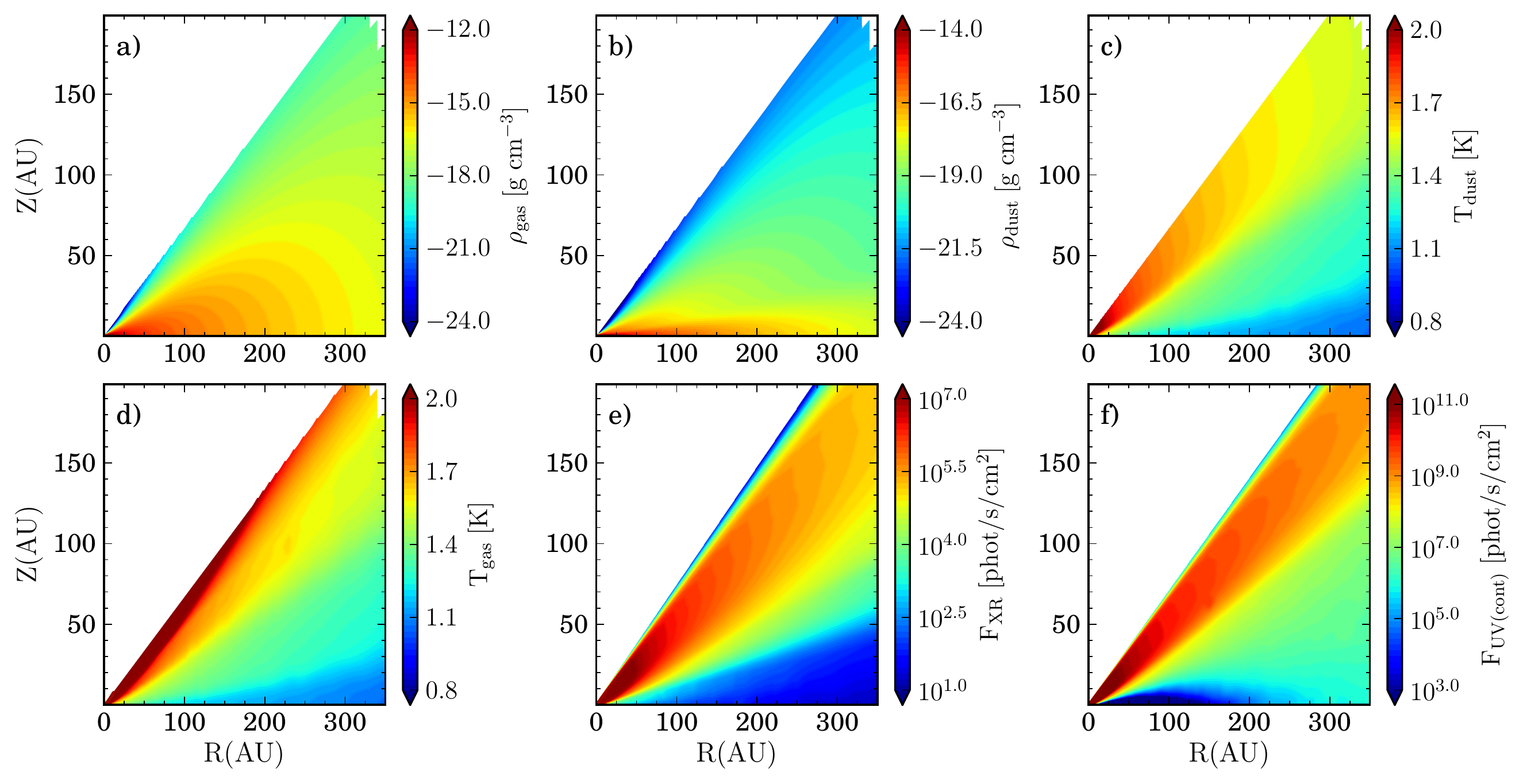}}
\caption{Figure illustrating the adopted disk physical structure for the kinetic model as a function of radius (R) and height (Z).  This
includes, the gas density (a), dust density (b),
dust temperature (c), gas temperature (d), UV photon flux (e), and X-ray photon flux (f).
Figure taken from Cleeves et al.\cite{cleeves13a}.  \label{fig:s4}}
\end{center}
\end{figure*}

\begin{table}[h]
\caption{Compilation of C and N Content in Solar System Bodies}
\begin{tabular}{@{\vrule height 10.5pt depth4pt  width0pt}lllll}
&\multicolumn3c{Relative Atomic Ratio\tablenote{See text for discussion of the upper and lower bounds, which represent ranges in measurements and/or uncertainties in a compiled measurement.}}\\
\noalign{\vskip-11pt}
\\
\cline{2-4}
\vrule depth 6pt width 0pt Source&\multicolumn1c{C/Si}&N/Si&C/N&Notes on CN Ratio\\
\hline
\multicolumn5c{Terrestrial Planets}\\
\hline
Earth (BSE) & $1.1(-3)\substack{+2.0(-4) \\ -2.0(-4)}$ & $2.5(-5)\substack{+5.0(-6) \\ -5.0(-6)}$ & 49.0 $\pm 9.3$ & This Text\tablenote{$1.0(-4)$ corresponds to $1.0 \times 10^{-4}$} \\
Earth Surface & $2.5(-4)\substack{+2.2(-5) \\ -2.2(-5)}$ & $1.5(-5)\substack{+2.7(-6) \\ -2.7(-6)}$ & 14.3 $\pm 2.9$ & This Text$^{\dagger}$\\ 
Earth Mantle & $8.7(-4)\substack{+2.2(-4) \\ -2.2(-4)}$ & $9.2(-6)\substack{+4.6(-6) \\ -4.6(-6)}$ & 80 $\pm 35$ & This Text$^{\dagger}$\\
Venus Atm. & $ 3.8(-4) $ & $ 2.8(-5) $ & 13.5 $\pm 2$ & \cite{Halliday2013}$^{\dagger}$\\
Mars Atm. & $ 1.6(-7) $ & $ 8.8(-9) $ & 18.0 $\pm 2$ & \cite{Halliday2013}$^{\dagger}$\\
\hline
\multicolumn5c{Meteorites\tablenote{As noted in the discussion regarding meteorites, the C/N ratio in carbonaceous chondrites was determined from different references than the C/Si and N/Si. However the measurements are consistent with the given upper and lower bounds.}}\\
\hline
CI & $0.71\substack{+0.14 \\ -0.50}$ & $0.042\substack{+0.011 \\ -0.011}$ &  $17.0 \pm 3.0$ & CI1\cite{Alexander13}\\
CO &  $0.07\substack{+0.11 \\ -0.02}$ & $0.003\substack{+0.002 \\ -0.002}$ &  $12.2 \pm 3.2$ & CO3 \cite{Alexander14}\\
CM & $0.40\substack{+0.08 \\ -0.21}$ & $0.022\substack{+0.010 \\ -0.015}$ &  $21.7\pm 1.3$ & CM2\cite{Alexander13}\\
CV &$0.08\substack{+0.14 \\ -0.04}$ & $0.004\substack{+0.005 \\ -0.003}$ &  $20.8\pm 3.7$ & CV3\cite{Pearson06}\\
L &$0.07\substack{+0.04 \\ -0.04}$ & $0.001\substack{+0.002 \\ -0.001}$ &  $45.2\pm25.2$ & L3\cite{grady89, Sugiura92, Sugiura98, hashizume95}\\ 
LL &$0.07\substack{+0.04 \\ -0.04}$ & $0.001\substack{+0.002 \\ -0.001}$ &  $56.0\pm43.7$ & LL3,5\cite{grady89, Sugiura92, hashizume95}\\
H & $0.04\substack{+0.01 \\ -0.01}$ & $0.00013\substack{+0.0001 \\ -0.00008}$ &  $452.0\pm298.0$ & H3,4\cite{kc78, grady89, Sugiura98}\\
EH & $0.022\substack{+0.008 \\ -0.008}$ & $0.002\substack{+0.001 \\ -0.001}$ &  $13.7\pm12.1$ & EH5,6\cite{gw03}\\
\hline
\multicolumn5c{Comets}\\
\hline
Comet Halley (dust) & $5.7\substack{+11.4 \\ -2.9}$ & $1.0\substack{+1.0 \\ -0.5}$ &  ..... & \cite{Delsemme91, wte91, Fomenkova99}\\
%Comet Halley (gas) & $2.6\substack{+0.5\\ -1.4}$ & $1.1\substack{+1.1 \\ -0.5}$ &  ..... & \cite{Delsemme91, wte91}\\
Comet Halley (dust + gas) & $8.3\substack{+8.3\\ -4.2}$ & $1.1\substack{+2.1 \\ -0.5}$ &  7.5$\pm$3.75\tablenote{For Comet Halley the errors on the C/Si and N/Si ratios are dominated by our assumed factor of 2 error on the Si measurement. For the C/N ratio we assume a 50\% error.  } & \cite{Fomenkova99}\\
Comet C/2011 W3  & $0.21\substack{+0.10\\ -0.10}$ & $0.03\substack{+0.01 \\ -0.01}$ & 6.2$\pm$0.9 & \cite{McCauley13}\tablenote{Nitrogen abundances relative to Si and C/N ratio provided by J. Raymond and given here.}\\
Comet C/2003 K7  & $0.08\substack{+0.04\\ -0.04}$ & .... & .... &\cite{McCauley13}\\
\hline
\multicolumn5c{Interstellar Medium}\\
\hline
ISM (Gas) & $41.7\substack{+8.0\\ -8.0}$ & $21.9\substack{+3.4 \\ -3.4}$ & 1.9$\pm$0.3 & \cite{Jensen07, Parvathi12}\\
ISM (Dust) & $5.85\substack{+1.5\\ -1.5}$ & $< 0.33$ & $> 17.57$ & \cite{Jensen07, Parvathi12, Jones13, Chiar13}\\
\hline
\end{tabular}
\end{table}

\end{document}